\documentclass[aps,preprint,amsmath,amssymb,superscriptaddress]{revtex4-2}
\usepackage{times}
\usepackage{graphicx}
\usepackage{dcolumn}
\usepackage{bm}
\usepackage{comment}
\usepackage{lipsum}
\usepackage{hyperref}
\usepackage{color}
\usepackage{epsfig}
\usepackage{latexsym}
\usepackage{mathtools}
\usepackage{caption}
\usepackage{subcaption}
\usepackage{bbold}
\usepackage{ulem}
\usepackage{verbatim}
\usepackage{url}
\usepackage{amsmath}
\usepackage[version=4]{mhchem}
\usepackage{gensymb}
\usepackage{textcomp}
\begin{document}

\title{Mechanical Clustering of Cells via the ExtraCellular Matrix}

\author{Ran Glinowiecki}
\affiliation{School of Mechanical Engineering, Tel Aviv University, Tel Aviv 69978, Israel}

\author{Shahar Goren}
\affiliation{School of Mechanical Engineering, Tel Aviv University, Tel Aviv 69978, Israel}
\affiliation{School of Chemistry, Tel Aviv University, Tel Aviv 69978, Israel}

\author{Oren Tchaicheeyan}
\affiliation{School of Mechanical Engineering, Tel Aviv University, Tel Aviv 69978, Israel}

\author{Bar Ergaz}
\affiliation{School of Mechanical Engineering, Tel Aviv University, Tel Aviv 69978, Israel}

\author{Robin L. B. Selinger}
\affiliation{Advanced Materials and Liquid Crystal Institute, Kent State University, Kent, Ohio 44242, USA}
\affiliation{Physics Department, Kent State University, Kent, Ohio 44242, USA}

\author{Yair Shokef}
\affiliation{School of Mechanical Engineering, Tel Aviv University, Tel Aviv 69978, Israel}
\affiliation{School of Physics and Astronomy, Tel Aviv University, Tel Aviv 69978, Israel}
\affiliation{Center for Physics and Chemistry of Living Systems, Tel Aviv University, Tel Aviv 69978, Israel}
\affiliation{Center for Computational Molecular and Materials Science, Tel Aviv University, Tel Aviv 69978, Israel}
\affiliation{International Institute for Sustainability with Knotted Chiral Meta Matter (WPI-SKCM$^2$), Hiroshima University, Higashi-Hiroshima, Hiroshima 739-8531, Japan}

\author{Ayelet Lesman}
\affiliation{School of Mechanical Engineering, Tel Aviv University, Tel Aviv 69978, Israel}
\affiliation{Center for Physics and Chemistry of Living Systems, Tel Aviv University, Tel Aviv 69978, Israel}
\email{ayeletlesman@tauex.tau.ac.il}

\begin{abstract}

Tissues are composites of living cells and extracellular matrix (ECM), that jointly determine their unique mechanical behavior. Here, we experimentally demonstrate that contractile cells remodel the ECM by generating long-range bands of aligned and densified fibers that mechanically couple cells into multicellular clusters. Using a finite-element model, we quantify the collective mechanical interactions mediated by these ECM bands for various volume fractions of cells. The model incorporates contractile particles that mimic cell-active forces within a nonlinear biological gel, inducing stiffening internally rather than through external loading. We find that even a small volume fraction of contractile particles (approximately $10\%$), much below what is expected for percolation of sphere contacts, leads to internal stiffening, mediated by the high-concentrated stress bands between cells and is strongly dependent on the nonlinear mechanical response of the ECM. Percolation analysis of the band network reveals sharp transitions at a critical particle volume fraction, that aligns with the onset of internal stiffening. These results demonstrate that long-range interaction between contractile cells through the deformation of the medium has the potential to globally shape the structure and mechanics of the bulk gel, leading to a functional structure with modified properties. These modified macroscopic changes can define organizing principles in tissue patterning and morphogenesis.

\end{abstract}

\maketitle

\newpage

\section{Introduction}
\label{sec:introduction}

The extracellular matrix (ECM) is composed of fibrous biopolymers, primarily collagen in loose connective tissues and fibrin in healing processes~\cite{Licup2015, Litvinov2017}. The fibrous structure of the ECM imparts unique non-linear elastic properties~\cite{grekas2021,  vanoosten2016, Litvinov2017, Notbohm-Lesman-Interface-2015, sarkar2024, piechocka2010structural, burla2019}; ECM fibers bend or buckle in compression and tense and align in tension, leading to unique behaviors such as softening in compression and stiffening in tension~\cite{shivers2020, song2021, xu-PRE-2015, Notbohm-Lesman-Interface-2015}. However, when the ECM is embedded with rigid particles that represent cells, the mechanical behavior of the ECM changes. Instead of causing buckling, external compression generates tension in the regions between the particles, leading to stiffening of the particle-gel composite~\cite{vanoosten2019, shivers2020, carroll2023, gandikota2020}. This type of behavior is reminiscent of the mechanical response of biological tissues~\cite{vanoosten2019}, thus allowing to represent complex living tissue with a rather minimal model that uses passive particles instead of active ones. Although these previous model systems have provided invaluable insight into the complex mechanics of biological materials, they lack an essential active component -- the intrinsic contractile activity that is characteristic of nearly all adherent cells~\cite{jansen2013}. 

Active cell contraction transmits forces to ECM fibers, reorganizing them into localized regions of alignment and compaction over distances that exceed cell size~\cite{hall2016, natan2020, goren2023, Goren-Koren-Xu-Lesman-2020, han2018cell}. The mechanism of this ECM remodeling can be described as localized strain-induced nematic ordering of the initially isotropic fiber network, and this disorder-to-order transition induces mechanical stiffening along the direction of fiber alignment~\cite{Yang2025, vader2009, grekas2021, Haimov2026}. The long-range ECM remodeling, generated by cell contractile forces, gives rise to extended ECM bands that mechanically couple neighboring cells and enables long-range interaction~\cite{natan2020, chen2022, nahum2023, alisafaei2021, sopher2018, wang2014, sawhney2002slow}. Interaction of cells via ECM bands was shown to coordinate various biological processes, including tissue morphogenesis~\cite{sawhney2002slow}, fibrosis~\cite{liu_matrix-transmitted_2020}, vascular assembly and capillary sprouting~\cite{korff_tensional_1999, Davidson2024, guo_long-range_2012}, tissue folding~\cite{hughes2018}, and cancer invasion and metastasis~\cite{shi_rapid_2014, kim_stress-induced_2017, ban2018}.

When many cells exert contractile forces on the ECM, aligned and mechanically stiff ECM bands form between neighboring cells, that can connect them into multicellular clusters or networks~\cite{natan2020, doha2022}. Fibroblasts seeded in fibrin gels connect into a large cellular network through such ECM bands. The cells that were part of the connected network exhibited a more spread morphology, whereas isolated cells remained more rounded and less spread, demonstrating that cellular connectivity through the ECM influences cell activity~\cite{natan2020}. The emergence of these three-dimensional cell clusters may have important consequences at multiple scales. At the cellular level, mechanical coupling through the ECM can transform individual cells into a coordinated collective, enabling cooperative behaviors that are not accessible to isolated cells. At the tissue scale, the interconnected network of mechanically reinforced ECM bands may alter the global mechanical properties of the cell-gel composite. From a physical perspective, this system can be viewed as a mechanically percolating network whose connectivity depends on cell density, similar to what was suggested for percolating actin-myosin networks ~\cite{kumar2023}. As the density of cells increases, the probability of forming a connected band with a neighboring cell rises, potentially giving rise to a percolation transition. Above the percolation threshold, a giant connected cluster emerges, enabling the connectivity to propagate throughout the system and leading to a dramatic change in its global behavior. In cell-ECM systems, several studies have recently begun to reveal the emergence of critical transitions and collective behavior. Experimentally, it was shown that fibroblasts embedded in fibrin gels induce rapid global gel contraction once cell density exceeds a critical threshold~\cite{doha2022, fernandez2009}. These experimental observations were further supported by computational studies~\cite{peng2025fiber, nan2019absorbing, kumar2026cooperative}. These findings suggest that mechanical interactions between cells, mediated by the ECM, may undergo collective transitions analogous to those observed in classical percolating elastic systems. Indeed, percolation of the cell-cell adhesion connectivity network has been shown to predict rigidity transitions in developing embryonic tissues~\cite{Petridou2021}. Yet, to the best of our knowledge, percolation theory has not been directly or systematically applied to the study of mechanical interactions between cells mediated through the ECM.

Here, we experimentally show that arrays of contractile cells can mechanically connect via ECM bands, forming multicellular clusters. We develop a computational model to explore the mechanical connectivity of such clusters at varying cell densities and their impact on global mechanical properties. In the model, we simulate contraction of particles (mimicking active cell contraction) that are embedded and coupled to a fibrin gel, represented by a hyperelastic continuum, and probed in shear to characterize its global mechanical properties. Our simulations reveal a notable internal stiffening effect at relatively low volume fractions of contractile particles, well below what is expected from effective medium theory or from percolation of mechanical contacts of spheres. Analysis of the resulting band networks uncovers a percolation transition that coincides with the onset of the internal stiffening. Our study demonstrates that long-range mechanical interactions between cells mediated by a mechanically nonlinear matrix can give rise to a percolation transition, leading to formation of a mechanically connected network that stiffens the material, a key factor in cancer and fibrosis.

\section{Cells Generate ECM Bands that Mechanically Couple Multiple Cells into Clusters}

\begin{figure}[t]
\centering
\includegraphics[width=0.8\textwidth]{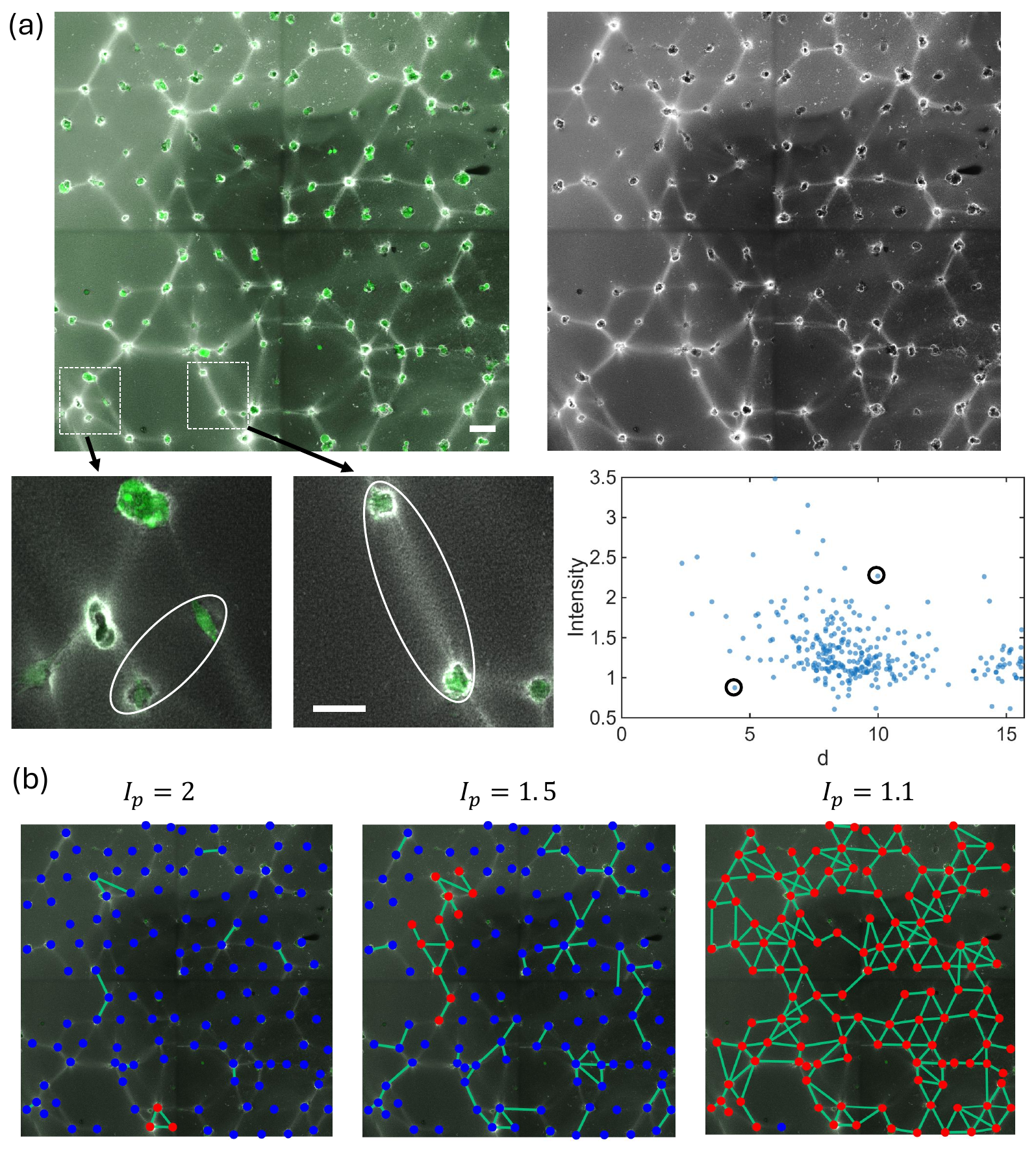}
\caption{(a) Micropatterned cancer cell aggregates (green) embedded on fibrin gel (white), generate deformed bands that connect multiple cell aggregates. Scale bar is $100\mu m$. White ellipses show two pairs of cell aggregates: one closely spaced pair with low ECM band intensity and one more distant pair with high ECM band intensity. Scale bar is $50\mu m$. The graph quantifies for all pairs the normalized fibrin intensity vs. intercellular distance, normalized by a typical cell radius of $13 \mu m$, as an approximation of the resultant cell aggregate radii~\cite{ergaz2024}. The two specific pairs are marked with black circles. (b) Analyzed clusters of cell aggregates by intensity threshold $I_p$. Band connections are marked in green, and the largest cluster is highlighted in red.}
\label{fig:exp_and_scheme}
\end{figure}

We used a previously developed micropatterning method~\cite{ergaz2024} (see Appendix~\ref{sec:exp-methods} for details) to organize cancer cell aggregates (HRAS-transfected epithelial cells labeled with GFP) into hexagonal arrays on a fluorescently-labeled fibrin gel (Fig.~\ref{fig:exp_and_scheme}a). Each cell-aggregate is about $60\mu m$ in diameter, and the typical distance between aggregates is $150\mu m$. Over several hours, cells apply contractile forces on the fibrin gel, generating extended bands of aligned and dense fibers between cell aggregates (Fig.~\ref{fig:exp_and_scheme}a). These bands are much stiffer than the background gel, and cells tend to migrate along the aligned fiber bands towards their neighboring aggregates~\cite{Goren2024,ergaz2024}. 

We quantified band intensity using the fluorescence labeling of the fibers. As shown in Fig.~\ref{fig:exp_and_scheme}a, the fluorescence signal (reflecting fiber density) exhibits large variation. Notably, band intensity is not consistently correlated with the distance between neighboring cell aggregates. For example, some pairs of cells at short separations do not exhibit visible bands, whereas other pairs at larger separations form pronounced bands. These observations indicate that intercellular distance alone does not uniquely determine band formation.

These bands mechanically connect cells into multicellular clusters. We identify such clusters by connecting pairs of cell aggregates with the gel intensity between them exceeding a threshold~$I_p$ (Fig.~\ref{fig:exp_and_scheme}b). As this threshold decreases, the largest connected cluster percolates the system in a manner reminiscent of a percolation transition~\cite{stauffer1994}.  

\begin{figure}[htbp]
\centering
\includegraphics[width=0.8\textwidth]{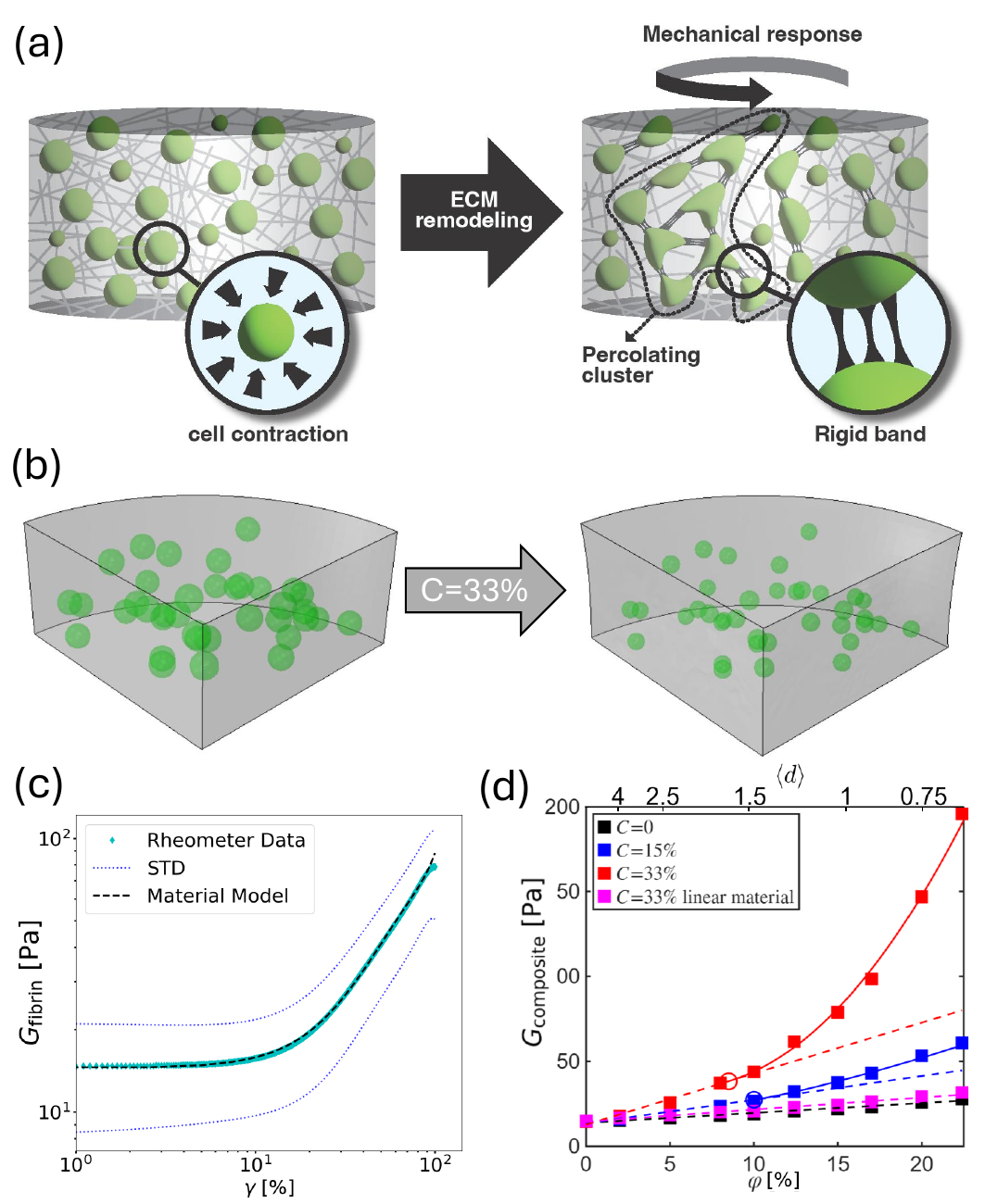}
\caption{ (a) Hypothesis of contracting cells altering the fibrous structure, forming bands that percolate and stiffen the global material. (b) Finite element model, before (left) and after (right) contraction by $C=33\%$. (c) Fibrin shear modulus to shear strain curve with the matching hyperelastic material model. (d) Composite shear modulus measured for increasing particle volume fractions, and for several contraction magnitudes as indicated in the legend. The linear material model at $33\%$ contraction is shown for comparison. Top axis shows the average inter-particle distance normalized by particle radius. Dashed lines show linear fitting to identify the beginning of nonlinear stiffening (open circles) and solid lines show power law fits beyond the critical density.}
\label{fig:material_properties}
\end{figure}

We anticipate that with increasing cell density, initially disconnected clusters will merge into larger, system-spanning structures, a hallmark of a percolation transition~\cite{Silverberg2014}. We hypothesize that such a percolation transition impacts the macroscopic mechanical behavior of the cell–gel composite, as schematically illustrated in Fig.~\ref{fig:material_properties}a. In the following sections, we introduce a computational model designed to capture the mechanical clustering of cells by these ECM bands within a relevant biological gel. Using this model, we study the collective long-range interactions between cells connected by ECM bands. Specifically, we ask whether there exists a critical transition point in the mechanical connectivity between cells that governs global mechanical changes.

\section{Establishing a Computational Model of Contractile Particles in Fibrin Gel}

Inspired by our cellular experiments (Fig.~\ref{fig:exp_and_scheme}), we selected fibrin as our model system due to its biological relevance as a fibrous material that mediates interactions between contractile cells, such as fibroblasts during wound closure. In this process, cells collectively engage to orchestrate tissue repair~\cite{geer2002}. Accordingly, we developed a computational model that simulates cell contractility represented by contracting particles embedded within a continuum with a material model of fibrin gel (Fig.~\ref{fig:material_properties}b). To accurately model fibrin in the simulations, we experimentally characterized the mechanical properties of fibrin gels at a concentration of $2\frac{mg}{ml}$ using a shear rheometer (Fig.~\ref{fig:material_properties}c). As expected, fibrin has a predominantly elastic response that is independent of frequency (see Appendix~\ref{sec:exp-methods})~\cite{zuidema2014, goren2023}. The linear shear modulus of fibrin is $14.7 \pm6.4 Pa$. At a strain of approximately $\gamma=10\%$, nonlinear stiffening emerges. We generated a continuum material model by fitting the experimentally measured rheology of fibrin gels using a hyperelastic polynomial strain energy density function, see Fig.~\ref{fig:material_properties}c and Appendix~\ref{sec:exp-methods}. 

In accordance with the rheometer experiment, we designed the continuum model to have the geometry of a cyclically symmetric quarter cylinder. Inside the bulk of the continuum fibrin cylinder, we embedded $100\mu m$ diameter rigid particles, mechanically linked to the gel, and distributed at random positions and with increasing volume fractions (see Appendix~\ref{sec:FEM}). We then simulated isotropic contraction of the particles, where we impose a contraction of the radius of each sphere from $r$ to $(1-C)r$, where $C$ quantifies the contraction. Contraction preserved the spherical shape, and the particles were not fixed in place. Depending on the volume fraction, between 15-170 contractile particles were embedded in the simulated domain.

\section{Composite Stiffness Increases with Volume Fraction of Contractile Particles}

We next evaluated the effect of particle contraction on the stiffness of the bulk composite. To measure the shear modulus of the contracted gel composite, we applied $5\%$ external shear strain on the bulk composite after the particles contracted. Because of internal prestress buildup in the material due to particle contraction, this is equivalent to a differential measurement, or a measurement of the tangent modulus~\cite{koenderink2009}. Increasing the volume fraction of the contractile particles led to bulk stiffening, initially exhibiting a linear dependence on volume fraction up to a critical volume fraction, beyond which a nonlinear power-law increase emerged. For contraction of $C=33\%$, stiffening is more profound, and non-linear stiffening begins earlier than for contraction of $C=15\%$, beyond approximately volume fraction of $\varphi_c=8.5\%$  and $\varphi_c=10\%$, respectively (end of linear fitting, dashed lines in Fig.~\ref{fig:material_properties}d). Contraction of $C=33\%$ led to an order of magnitude stiffer composite compared to a gel embedded with passive non-contractile particles (Fig.~\ref{fig:material_properties}d). Finally, contraction in a linear elastic material barely showed any stiffening even at high contraction of $C=33\%$ and was quite comparable to passive particles, namely $C=0\%$ (Fig.~\ref{fig:material_properties}d).

\section{Bands connect particles into networks}

\begin{figure}[htb]
\centering
\includegraphics[width=0.8\textwidth]{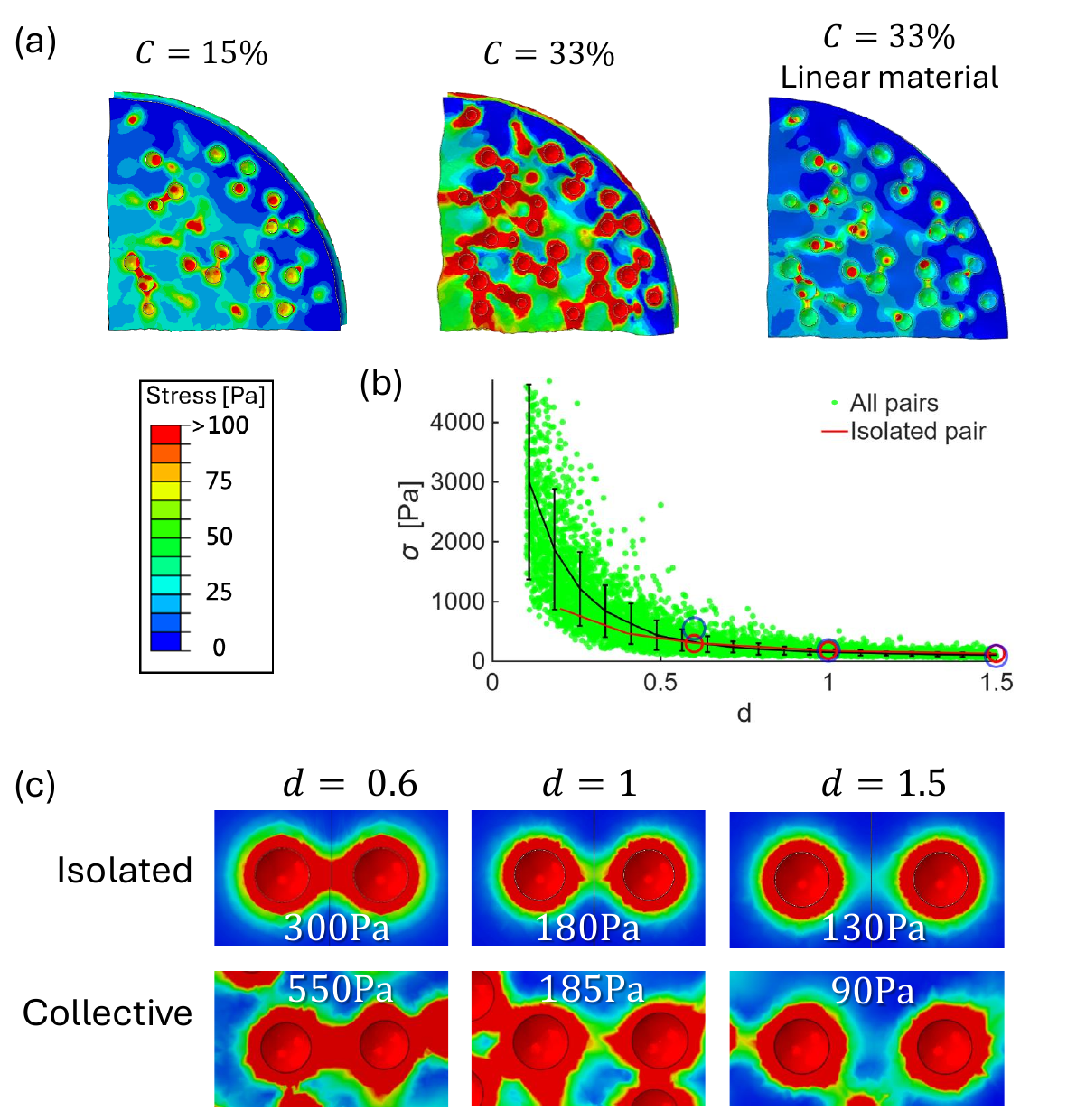}
\caption{High-stress bands form between contractile particles in the nonlinear simulated gels. (a) Finite element analysis with $\varphi = 17\%$ volume fraction of contracting particles. Images show top view cut section of von Mises stress, for contraction of $C=15\%$, $C=33\%$ (both for nonlinear material) and $C=33\%$ with linear material properties. (b) Stress in bands versus pair distance (normalized by particle radius) for all pairs of particles at $C=33\%$ in the collective system (green cloud of points) with the average and standard deviation in black. Comparison with an isolated pair is shown in red. (c) Stress maps at $C=33\%$ of bands in isolated pairs and in the collective system at the same normalized distances. The value of average stress in the band is mentioned on each image. These pairs are marked in red (for isolated) and blue (for collective) dots in (b).}
\label{fig:cloud}
\end{figure}

To better understand the mechanical changes of the contractile particle-gel composite, we delve into the mechanical responses at the micro-scale resulting from the contracting particles. Figure~\ref{fig:cloud}a shows the internal stress distribution after the particles have contracted by $C=15\%$ and $33\%$. Supplementary Movie 1 shows the evolution of the developed strains and stresses during the contraction process. Increasing particle contraction leads to visible bands of concentrated stress between nearby particles. Changing the material model to a linear elastic material results in significantly lower stress, an order of magnitude smaller, and the visible band network that was evident in the nonlinear material is completely absent in the linear case. Interestingly, consistent with the experiments (Fig.~\ref{fig:exp_and_scheme}a), not all neighboring pairs appear to form bands, and the band intensity (i.e., the mean stress within a band) can vary substantially between different neighboring pairs.

To characterize the bands in the collective system of particles, we compare it with the minimal case of two contractile particles at increasing distances inside an otherwise empty gel (Fig.~\ref{fig:cloud}b,c). When the two particles are far from each other, the average stress in the region between them is relatively constant at $90-100Pa$ (for $C=33\%$), generated by the contraction of each isolated particle. However, when the particles are closer together, the stress begins to concentrate in the region between them, forming a band. We then compared the stress-distance relationship of the isolated pair with the collective particle system. For the collective system, we analyzed the average stress in bands between neighboring pairs across all simulated gels with varying particle volume fractions. Figure~\ref{fig:cloud}b plots the band-stress over all pairs at every given distance. It can be seen that the collective system exhibits a broad distribution of stress values (shown as a green cloud of dots), with differences in stress of up to an order of magnitude for pairs at the same distance. This means that bands forming between particles at the same distance can have highly variable stress values, often significantly deviating from those observed in isolated pairs. This indicates that distance alone does not uniquely determine stress. As an example, we selected several pairs from the collective system and compared them with isolated pairs positioned at the same distance (Fig.~\ref{fig:cloud}c). Notable differences in band stress are observed between these systems. This variability cannot be explained by pairwise interactions alone and instead points to cooperative effects arising from the mechanical influence of surrounding particles.

\section{Percolation Analysis of the Band Network}

\begin{figure}[htb]
\centering
\includegraphics[width=\textwidth]{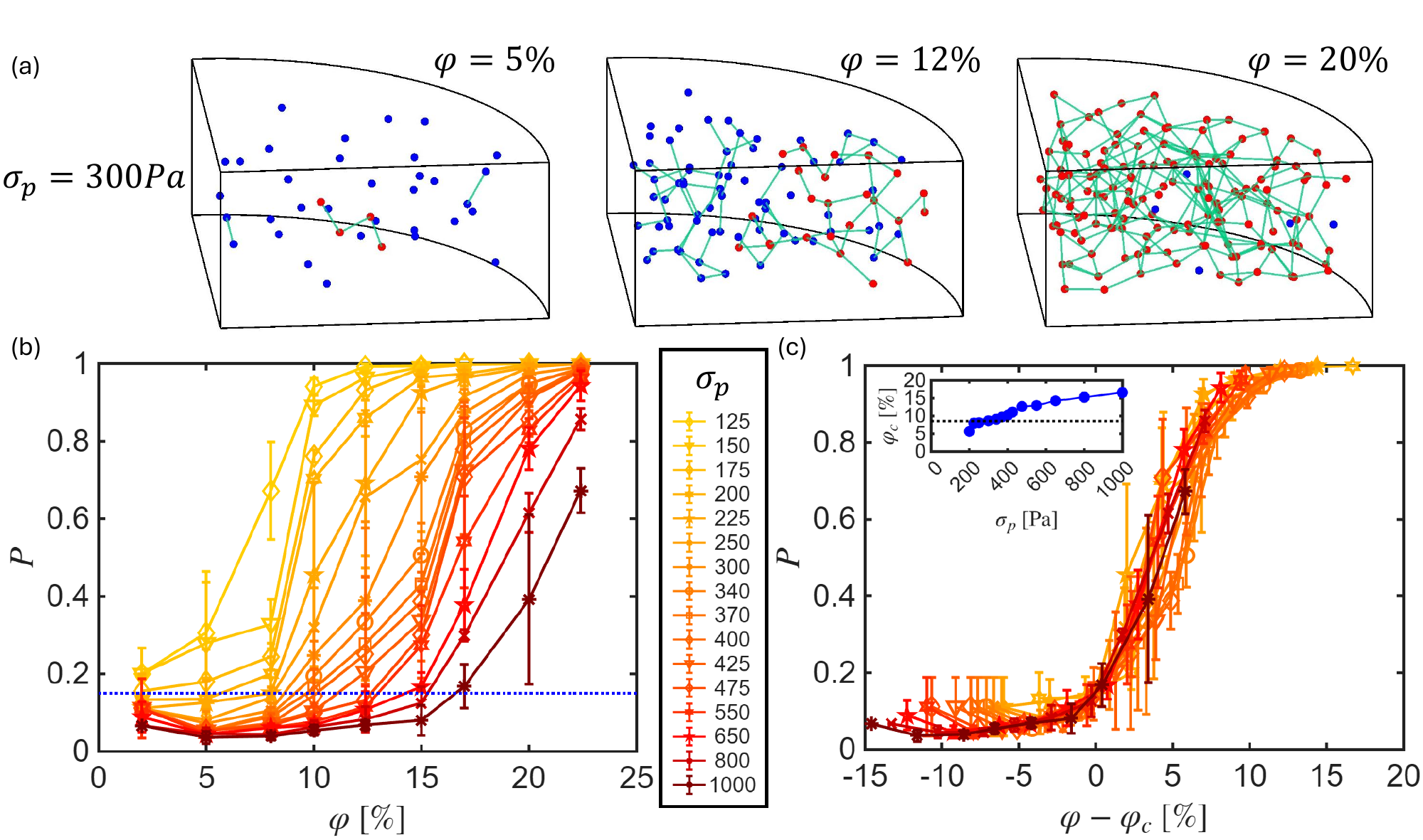}
\caption{Percolation analysis for sphere contraction of $C=33\%$. (a) Band networks in the simulations, at different volume fractions, using a stress threshold of $\sigma_p = 300 Pa$. The positions of the spheres are marked with blue dots, bonds exceeding the stress threshold are green, and spheres belonging to the largest cluster are marked in red. (b) The strength of the largest cluster as function of sphere volume fraction for a wide range of stress thresholds, as indicated in the legend. (c) Strength of the largest cluster as function of the volume fraction shifted with respect to the critical volume fraction (defined by crossing $P=0.15$, marked with a blue dotted line in panel b). Inset: the dependence of the critical volume fraction on stress threshold. The dotted black line shows the case for stress threshold of $300 Pa$ chosen for panel a.}
\label{fig:percolation}
\end{figure}

We characterized the band network using a clustering analysis, similar to that applied to the experimental data shown in Fig.~\ref{fig:exp_and_scheme}b. In the simulations, the bands act as bonds and the spheres act as nodes. We selected a mechanical threshold $\sigma_p$, such that when the average stress between two particles is higher than the threshold, a bond between them is positioned~\cite{ostojic2006, pastor-satorras2012}. The threshold values were tested over a wide range of possible values. In Fig.~\ref{fig:percolation}a, we show the identified networks of spheres connected by mechanical bonds for a mechanical threshold of $\sigma_p = 300 Pa$ and for an increasing volume fraction of particles. This threshold captures the high-stress bands between cell pairs, high enough to exclude background stress ($<100 Pa$), yet not too high that genuine bands are lost. The spheres that are part of the largest connected cluster are highlighted in red.  

The strength $P$ of the largest cluster was quantified by normalizing the number of spheres in the largest cluster by the total number of spheres in the system. We then plotted $P$ as function of volume fraction for different thresholds, see Fig.~\ref{fig:percolation}b, where each curve corresponds to a different stress threshold $\sigma_p$. For all thresholds, a sharp increase in the strength of the largest cluster within a narrow range of densities is observed, indicating a percolation transition. Increasing the threshold shifts this transition toward a higher critical density. We define the critical volume fraction, $\varphi_c$, as the point where the strength of the largest cluster crosses the arbitrary value $P=0.15$ (horizontal blue line in Fig.~\ref{fig:percolation}b). This value corresponds to the onset of the sharp increase in cluster size for most stress thresholds. The critical volume fraction $\varphi_c$ as function of the threshold $\sigma_p$ is shown in the inset of Fig.~\ref{fig:percolation}c, showing that $\varphi_c$ is within the range of $5-15\%$. Finally, by subtracting $\varphi_c$ from the varying $\varphi$, we see that all curves collapse into a single unified behavior (Fig.~\ref{fig:percolation}c), indicating that the percolation behavior is non-threshold-dependent.

\section{Comparison with effective medium theory}

It is instructive to contrast our findings with known results from effective medium theory to calculate the properties of a composite, where percolation of sphere-sphere direct contacts plays a key role~\cite{lorenz_2001, snarskii2020, lefevre_EML_2022}. Simulation results of our model for embedded stiff spheres with no contraction (Fig.~\ref{fig:material_properties}d) show agreement with effective medium theory in the low density limit, where the effective shear modulus $G_{\rm composite}$  of the composite increases linearly with volume fraction of rigid spherical inclusions. While we did not study volume fractions above $22.4 \%$, for passive spheres, we expect a rapid increase of $G_{\rm composite}$ only at considerably higher volume fractions; percolation of overlapping spheres is expected at $\varphi_c=0.4$~\cite{lorenz_2001}, effective medium theory predicts from duality considerations a critical density $\varphi_c=0.5$~\cite{snarskii2020}, and random close packing occurs at $\varphi_c=0.64$~\cite{torquato_RMP_2010}. As we have shown in our simulations, contraction of the embedded spheres, combined with nonlinear properties of the host medium, enables formation of highly correlated stiff bands which act as force chains, at volume fractions far below the percolation threshold of the embedded sphere-sphere contact. Stress is concentrated on a sparse, correlated network of strain-stiffened nematic bands which induce strong nonaffine deformation. As a result, the effective shear modulus $G_{\rm composite}$ shows onset of nonlinear dependence at a significantly smaller volume fraction of inclusions at $8.5\%$ or $10\%$ (Fig.~\ref{fig:material_properties}d) compared with the predication of effective medium theory.

For sphere contraction of $C=33\%$, the critical volume fraction of $\varphi_c = 8.5 \%$ for the nonlinear increase in shear modulus $G_{\rm composite}$ coincidences with the range of 5-15\% volume fractions where band percolation transition occurs, as determined from our percolation analysis (Fig.~\ref{fig:percolation}c, inset). This suggests that the onset of percolation of the network of stiffened bands--not percolation of contacts between the embedded spheres--triggers sharp increase of the effective shear modulus of the composite. 

Above the percolation threshold, we expect scaling behavior for the composite's shear modulus, $G_{\rm composite} \sim (\varphi-\varphi_c)^f$ with a scaling exponent $f \approx 3.75$~\cite{roux_relation_1986, sahimi_relation_1986, wilbrink_rigidity_2005}. After subtracting the low-density linear scaling $G_{\rm lin}=A+B\varphi$, we find power law scaling $G_{\rm composite} - G_{\rm lin} \propto (\varphi-\varphi_c)^f$, however with $f=2$ for a contraction level of $C=33\%$ and $f=1.5$ for $C=15\%$ (see solid lines in Fig.~\ref{fig:material_properties}d and more details in Appendix~\ref{app:power-law}). Clearly the system that we study is different than the simpler percolation scenario of passive inclusions, yet it is interesting that the exponents that we obtain differ and vary with contraction.

\section{Discussion}

The simulations presented in this work are largely motivated by our cellular experiments showing that arrays of fibroblast aggregates establish multiple mechanical connections through the deformation bands they generate in the matrix. The mechanical links formed between cells give rise to a clustering effect. Our model consists of contractile particles embedded within a nonlinear material, thus capturing the active contraction of cells in the ECM. Particle contraction results in the formation of deformed bands of concentrated stress that extend between neighboring particles, clustering cells together, similar to the effect observed in the experiments. 

Our model offers deeper insight into the network of interaction that cells establish through the deformations they impose on the ECM~\cite{Shokef-Safran-PRL-2012, golkov-NJP-2017, golkov-EPJE-2024, zemel2006, ben-yaakov2015}. In particular, it elucidates the relationship between local, cell-scale mechanical interactions and the emergence of global connectivity and mechanical transitions through percolation of the cell-induced band network. Our work builds on important prior studies showing a phase-transition–like behavior arising from collective cellular interactions~\cite{doha2022, peng2025fiber}, and showing cooperative effect of local active stresses on the macroscopic contractility of elastic fiber networks~\cite{kumar2026cooperative}. 

A possible experimental system to compare with our model could be using PANIPAAm (Poly(N-isopropylacrylamide)) particles in fibrin or collagen~\cite{sarkar2024}. By heating the PNIPAAm particles at their lower critical solution temperature (LCST) they contract. Their dimensions and contractility are similar to those used in our work~\cite{grekas2021, burkel2017}. The LCST of PNIPAAm is close to the body temperature, making an intriguing possible future synthetic tissue composite that can stiffen internally by heating as an external signal, and the stiffness of the composite can be controlled by the relations between volume fraction and temperature.

Our study is relevant to biological processes such as wound healing or fibrosis, in which (myo-) fibroblasts exert forces on the ECM fibrous matrix and, over time, their contraction drives the collective healing process. In the early stages of healing, cells are more sparsely distributed and may not be in direct contact, making their ability to interact mechanically through the ECM essential for coordinated, collective wound closure. Our study suggests that the network of interaction between cells and percolation of that network is critical in mediating the global mechanical transitions, such as global contraction or stiffening of the tissue.

\section*{Acknowledgments}
We thank Fred MacKintosh, Tomer Markovich, and Jonathan Selinger for helpful discussions. This work was partially supported by Grant No. 2022197 from the United States-Israel Binational Science Foundation. Ayelet Lesman acknowledges support by the Israel Science Foundation (Grant No. 2072/23).

\appendix

\section{Experimental Methods}
\label{sec:exp-methods}

\subsection{Fibrin Gel Preparation}

Fibrin gel at a final concentration of $2\frac{mg}{mL}$ was prepared by mixing human fibrinogen $4\frac{mg}{mL}$ (Evicel Biopharmaceuticals) with human thrombin $2\frac{U}{mL}$ (Evicel Biopharmaceuticals) at a ratio of 1:1. Fibrinogen concentration was chosen to be within the physiological range~\cite{vilar2020} and a concentration relevant for cell culturing~\cite{montero2021}. Thrombin concentration was chosen to allow sufficiently slow polymerization for setting up rheometer experiments. 

\subsection{Cell Patterning}

Cell experiments were preformed as previously described~\cite{ergaz2024}. In short, GFP-labeled murine epithelial cells harboring the cancerous HRAS mutation were cultured in DMEM (Dulbecco’s Modified Eagle Medium) supplemented with $10\%$ fetal bovine serum, nonessential amino acids, sodium pyruvate, L-glutamine, $100 \frac{U}{mL}$ penicillin, $100 \frac{\mu g}{mL}$ streptomycin, and $100 \frac{\mu g}{mL}$ neomycin. The cells were cultured in a $37 \degree C$ humid incubator. Patterned coverslips were created using photolithography using a commercial micropatterning kit (4DCell\textcopyright, Montreuil, France). First, $24 mm$ diameter coverslips were passivated with a poly-lysine Poly(ethylene glycol) (PEG) solution (PLL-g-PEG, $0.1 \frac{mg}{mL} $, from 4DCell\textcopyright) and incubated for 30 minutes. Next, the coverslips were rinsed with distilled water and placed on a customized photomask made of a quartz substrate with a thin chrome layer. The photomask and coverslips were stacked inside an aluminum frame and exposed to UV light ($220 nm$ wavelength) for 10 minutes using a ProCleaner\texttrademark{} Plus system. After UV exposure, we removed the patterned coverslips, rinsed them again, and prepared silicone rubber molds. These molds, cut from $0.5 mm$ thick polysiloxane (from McMaster-Carr), contained four circular cutouts (each $6 mm$ in diameter). The silicone was washed with $10\%$ sodium hypochlorite and sterilized in $70\%$ ethanol. Silicone molds were attached to the $24 mm$ patterned coverslip to serve as containers for cell suspension and gel formation. HRAS cancer cell suspensions (approximately $9 \cdot 10^3$ cells per silicone cutout) were added to each mold, followed by incubation. Unattached cells were washed away, and the samples were incubated for up to 48 hours in a $37 \degree C$ humid incubator. Finally, we added $14 \mu L$ of fluorescently labeled fibrinogen ($10 \frac{mg}{mL}$; Omrix Biopharmaceuticals) and gently mixed it with $ 14 \mu L$ of thrombin ($10 \frac{U}{mL}$; Omrix Biopharmaceuticals) in each silicone mold. The samples were incubated for 30 minutes to polymerize, and warm medium was added to cover the gels. Using tweezers, we gently lifted the micropatterned gel from the silicone.

\subsection{Microscopy}

Imaging was done 4.5 hours after lifting the gel with patterned cells from the coverslip, which allowed the cells to begin deforming the environment. The gels with patterned cells were imaged using a Zeiss LSM 880 confocal microscope equipped with an Airyscan fast detector. A 40× water immersion and 10× objective lenses (Zeiss) were used along with a $30 mW$ argon laser (at wavelengths $488$ and $514 nm$). During imaging, the cells were kept in a $37 \degree C$, $5\%$ CO$_2$ incubation chamber. 

\subsection{Image Analysis}

The patterned cells $z$ stack of images first underwent maximum intensity projection. Cell centers were identified using imageJ, with a constant cluster diameter of $20\mu m$~\cite{ergaz2024}. Using a custom made Matlab code, the average intensity value of the fibrin gel channel inside the selected band area was normalized to a manually selected area of background near the band, which allowed the use of an image with uneven background. The band dimensions were taken to be the same as was done for the simulation percolation analysis (width of one radius of cell aggregate). Clusters were identified in the same manner as they were in the percolation analysis, based on normalized inter-cellular distance versus normalized intensity. Pairs that have degraded gel, bad image stitching or interfering cells between them were discarded from the analysis.

\subsection{Bulk Rheology}
\label{sec:bulk-rheology}

\begin{figure}[h]
\centering
\includegraphics[width=\linewidth]{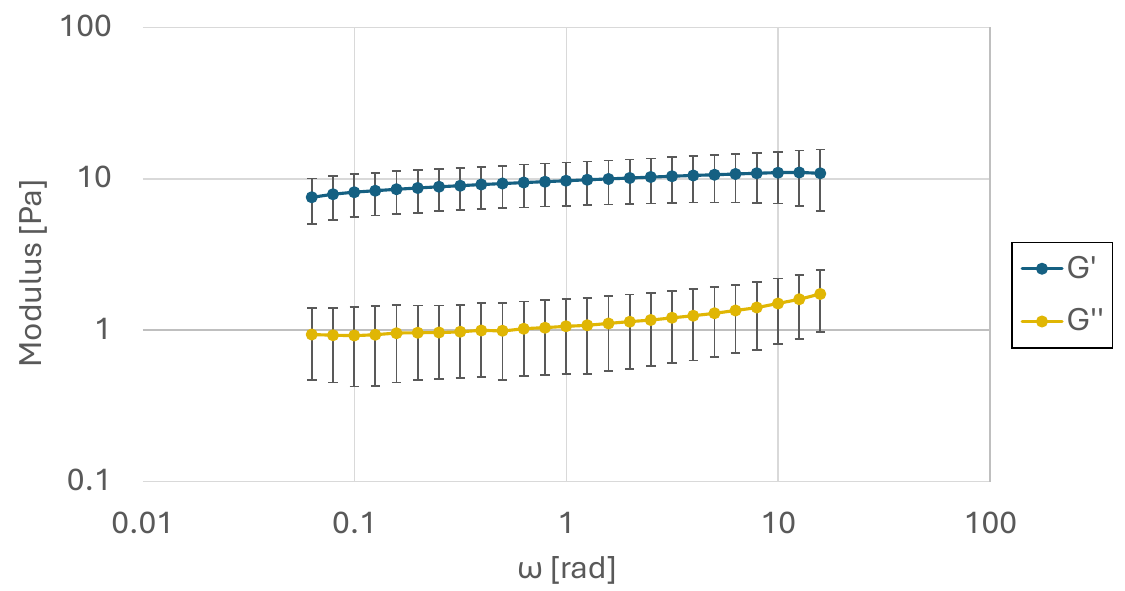}
\caption{Frequency sweep at $5\%$ strain, for $2\frac{mg}{mL}$ fibrin}
\label{fig:frequency}
\end{figure}

Rheology experiments were performed on a Discovery HR-3 rheometer using a $20 mm$ diameter parallel plate at a gap of $0.5-1 mm$. In order to allow sufficient time to polymerize the fibrin gel within the rheometer, the plate was cooled to $10^\circ C$ prior to pipetting the fibrinogen. Fibrinogen and thrombin were pipetted on the bottom plate at a 1:1 ratio. Total sample volume was $\sim170\mu L$ to fit the desired gap. Gel polymerization was initiated between the plates at 37$^\circ C$ for half an hour~\cite{zuidema2014}. A humid solvent trap was used to prevent drying of the sample and PBS was poured around the polymerized sample and left to rest for 15 minutes prior to starting the experiment. Strain amplitude measurements were done between $0.1-100\%$ strain~\cite{Hyun2011} at a constant frequency of $0.5~Hz$ and frequency sweeps were preformed between $0.02-20~Hz$ at a constant strain of $5\%$. Figure~\ref{fig:frequency} shows that the elastic storage modulus $G'$ was found to be much larger than the viscous loss modulus $G''$, and both hardly depend on frequency. Those measurements were preformed over seven samples. 

\section{Finite Element Simulations}
\label{sec:FEM}

\begin{figure}[h]
\centering
\includegraphics[width=\linewidth]{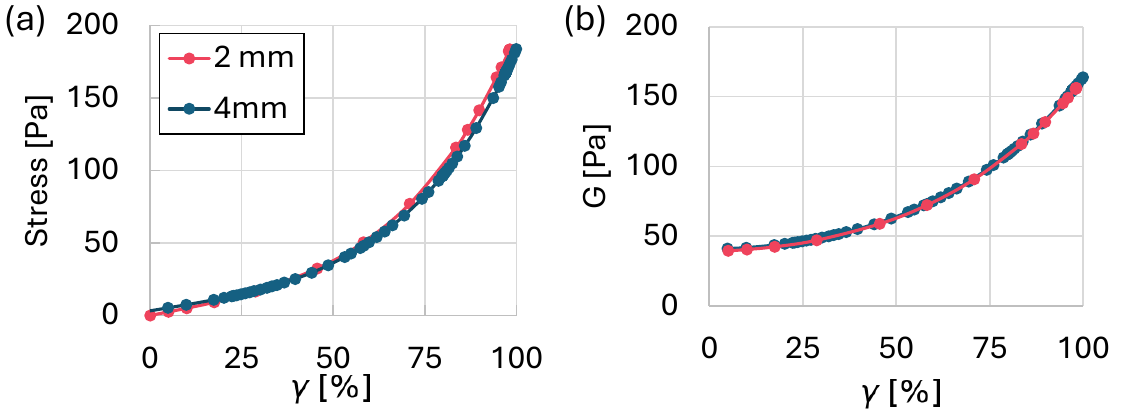}
\caption{Effect of composite diameter on output results in a $\varphi = 10\%$ model on (a) perimeter stress output, and (b) shear modulus.}
\label{fig:diameter}
\end{figure}

Finite element software Abaqus/CAE (Dassault Systemes Simulia, Johnston, RI) was used to simulate a fibrin-like continuum material embedded with contractile particles at various volume fractions. The dimensions and loading conditions were chosen to replicate a shear-loading rheometer experiment to enable direct comparison with the fibrin gel rheology (Fig.~\ref{fig:material_properties}c). Specifically, we attempted to replicate a parallel plate rheometer experiment that is typically used in rheology of biological gels, considering an $8 mm$ gel diameter (plate diameter) and $0.5 mm$ gel thickness (gap distance)~\cite{zuidema2014}. However, a full-sized $3D$ model could not be used due to extensive computational requirement, and therefore a reduced $3D$ model was implemented with a smaller gel diameter of $2 mm$. Nevertheless, in Fig.~\ref{fig:diameter}, we show that the difference in gel diameter did not significantly affect the obtained results. In addition, the sample was modeled as a quarter cylinder with cyclic symmetry. Evenly sized spherical particles $100\mu m$ in diameter were placed randomly within the cylinder, using a custom-made python code. The particles were not allowed to be positioned within a minimal surface to surface distance of $5\mu m$, in order to allow for a sufficient number of elements to be placed between them, to allow adequate calculations. The number of particles placed was determined by the desired volume fraction $\varphi$,
\begin{equation}
\varphi = \frac{V_{\rm particle} \cdot n_{\rm particle}}{V_{\rm gel}} , 
\label{eq:phi}
\end{equation}
where $V_{\rm particle}$ is the volume of a single sphere, $n_{\rm particle}$ is the number of spheres and $V_{\rm gel}$ is the volume of the gel.

\begin{figure}[h]
\centering
\includegraphics[width=\linewidth]{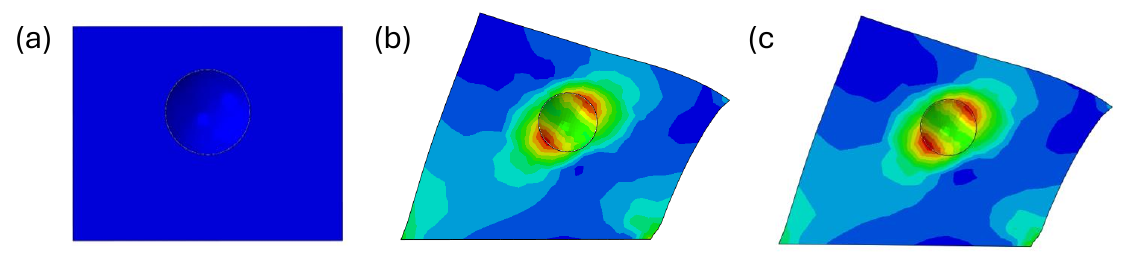}
\caption{Comparison of stresses in a contractile sphere ($C=25\%$) and kinematic coupling in a continuum, which is then sheared (the sphere is hidden). (a) before contraction and shear, (b) stiff sphere (hidden) and (c) kinematic coupling, both contracted and sheared. demonstrating the change of size, conservation of spherical shape (rigidity), and the ability of the sphere center to change position in both cases.}
\label{fig:kinematic_coupling}
\end{figure}

\begin{figure}[h]
\centering
\includegraphics[width=\linewidth]{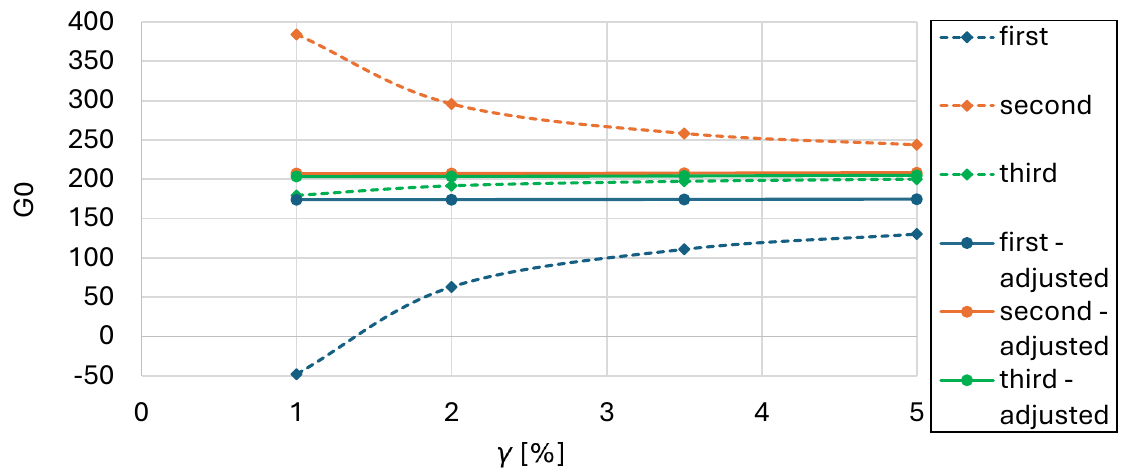}
\caption{(a) Three models at $\varphi=22.4\%$ showcasing the dependency of linear shear modulus $G_0$ to the external shear strain. Displaying both before and after adjusting the data to the zero-strain torque measured, as explained in the methods.}
\label{fig:modulus_linearity}
\end{figure}

The particles were then subtracted from the gel, making holes in their place. A kinematic coupling boundary condition was defined on the borders of the holes, instead of using the actual particles. The kinematic coupling is a type of exact constraint fixture constraining all 6 degrees of freedom relative to the sphere center, effectively defining them as rigid inclusions. The kinematic coupling also allowed for active contraction which mimics either the activity of live cells, or the contraction of PNIPAAm microgel spheres responding to lowering the temperature. The kinematic coupling boundary condition created a rigid sphere adhered to the gel. Its center was not fixed, allowing the sphere to move if pulled by the gel. We demonstrate that this methodology generates spheres that behaves the same as physical spheres without the need to directly compute them (Fig.~\ref{fig:kinematic_coupling}). The bottom plate was fixed, and the top plate was fixed to a $z$ height of $0.5 mm$, matching the boundary conditions of the rheometer experiment Fig.~\ref{fig:material_properties}c). The simulated spheres were contracted to the desired amount (usually a contraction of $C=33\%$, unless otherwise specified), followed by an external mechanical perturbation (usually shear strain of $\gamma=5\%$, unless otherwise specified). Superposing this small shear strain externally on the internally-stressed material, after the particles have already contracted, allows to measure the shear modulus $G$. Particle contraction applied torque in the system (before applying the external strain - zero strain torque).  This was offset by subtracting it from the torque measured after applying the external strain, as demonstrated in Fig.~\ref{fig:modulus_linearity}. In Fig.~\ref{fig:modulus_linearity} we show that the applied external strain used for probing the stiffness has no effect on the shear modulus, because at these small strains the composite behaves linearly.

\begin{figure}[h]
\centering
\includegraphics[width=\linewidth]{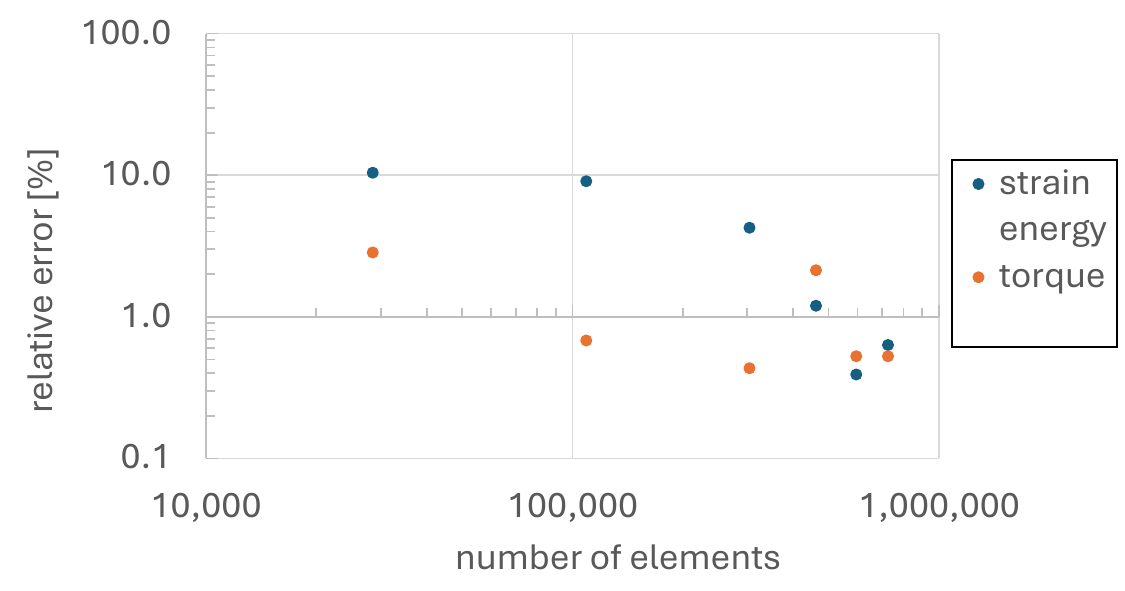}
\caption{Convergence test of strain energy and torque. Values are the relative error compared to finest mesh size of 850,000 elements.}
\label{fig:convergence}
\end{figure}

Meshing was done using adaptive mesh, which runs several iterations of the simulation and refines the mesh within mechanically sensitive regions. Element type was $C3D10H$, a 10-node quadratic tetrahedron, with hybrid formulation to deal with the material incompressibility. A convergence test was preformed compared to $\sim850,000$ elements, at which point values stopped changing drastically with mesh refinement (Fig.~\ref{fig:convergence}), and according to its results the number of elements used was approximately $500,000$ (measured values had relative errors of under $1\%$) with average computation time of about 8 hours on a standard desktop computer.

A  simulation of a pair of isolated particles was also preformed. A half sphere of fibrin continuum with radius of $1 mm$ was modeled, and plane symmetry was defined. A particle was placed in alternating distances. All parameters were identical to the other non isolated simulations.

The material model that best fitted our rheometer data of fibrin gel was a hyperelastic rank 4 reduced polynomial strain energy density function,

\begin{align}
U &= 7.226 \left( \tilde{I}-3 \right) + \frac{1}{4.194 \cdot 10^{-3} } \left(J-1\right)^2 + 33.543 \left(\tilde{I} - 3 \right)^2 \nonumber \\
&- 21.484 \left(\tilde{I} - 3 \right)^3 + 8.58 \left( \tilde{I} - 3 \right)^4 [10^{-3} \cdot \frac{J}{m^3}] ,
\label{eq:U_full}
\end{align}
where $\tilde{I}_1=I_1 \cdot J^{2/3}$ ,$I_1=\lambda_1^2+\lambda_2^2+\lambda_3^2$, as $\lambda_i$ is the principal stretch in the $i$ direction, and $J$ is the Jacobian of the deformation gradient. The material is approximately incompressible, with Poisson’s ratio defined as 0.485, the strain energy density function within the incompressibility approximation is,
\begin{equation}
U = 7.226 (I_1-3) + 33.543 (I_1-3)^2 - 21.484 (I_1-3)^3 + 8.58 (I_1-3)^4 ,
\label{eq:U_incompressible}
\end{equation}

The torque needed to strain the sample was used to calculate the linear shear modulus $G_0$ using the following formula,
\begin{equation}
G_0 = \frac{T \cdot h}{J_z \cdot \Delta \theta} ,
\label{eq:_G0}
\end{equation}
where $T$ is torque (removed zero strain torque), $h$ is height of the gel, $J_z$ is polar moment of inertia and $\Delta \theta$ is the rotational displacement.

\section{Characteristic Inter-Particle Distance and Distance Normalization}
\label{sec:distNorm}

For each volume fraction of particles, the characteristic inter-particle distance was calculated, by dividing the gel volume by the number of particles and taking the cubic root. To obtain the typical surface to surface distance between neighboring spheres we further subtract two particle radii,
\begin{equation}
d_{\rm characteristic} = \left( \frac{V_{\rm gel}}{n_{\rm particle}} \right)^\frac{1}{3} - 2 r_{\rm particle} = \left( \frac{V_{\rm particle}}{\varphi} \right)^\frac{1}{3} - 2 r_{\rm particle} ,
\label{eq:d_average}
\end{equation}
where in the second equality, we used Eq.~(\ref{eq:phi}) to rewrite in terms of volume fraction. Note that the particle radius used here is before contraction.

Normalized distance is calculated between any pair of particles, cells, or normalizing the characteristic average inter-particle distance calculated at Eq.~(\ref{eq:d_average}) (at Fig.~\ref{fig:material_properties}d), in the following manner:
\begin{equation}
d = \frac{D - 2 r_{\rm particle}}{r_{\rm particle}} ,  
\label{eq:d_r}
\end{equation}
when D is the distance between the two particles/cells in $\mu m$.

\section{Percolation Analysis}

Percolation analysis was done using a custom made Matlab code, creating a graph whose nodes are the particles as positioned in the finite element simulations, in $3D$ space. The links (or bands) are determined by measuring the average stress between all possible pairs and comparing it to a threshold stress, referred to as the mechanical threshold $\sigma _p$. The stress averaging is done by averaging the elemental stress data inside a cylinder between the two particles, with a diameter of the particle original (non contracted) radius. The links formations was only possible if there was no other particle present within the cylinder boundaries. The largest connected cluster was marked and its strength was calculated as the number of spheres in it divided by the total number of spheres in the system. Images displaying the created networks are presented in Fig.~\ref{fig:percolation}a, the spherical particles are depicted smaller than they actually are, to allow for better visualization of the bands. When looking at the critical largest cluster size, chosen to be $P=0.15$, at each threshold, we get its equivalent critical volume fraction $\varphi_c$. Ideally, with an infinite system size, we would expect $P=0$ for $\varphi<\varphi_c$ and then to start growing as $\varphi>\varphi_c$, but because of the finite size of the system we cannot get $P=0$, so a $P=0.15$ was chosen for assessing $\varphi_c$.

\section{Power-Law Scaling of Composite's Stiffness}
\label{app:power-law}

\begin{figure}[h]
\centering
\includegraphics[width=\linewidth]{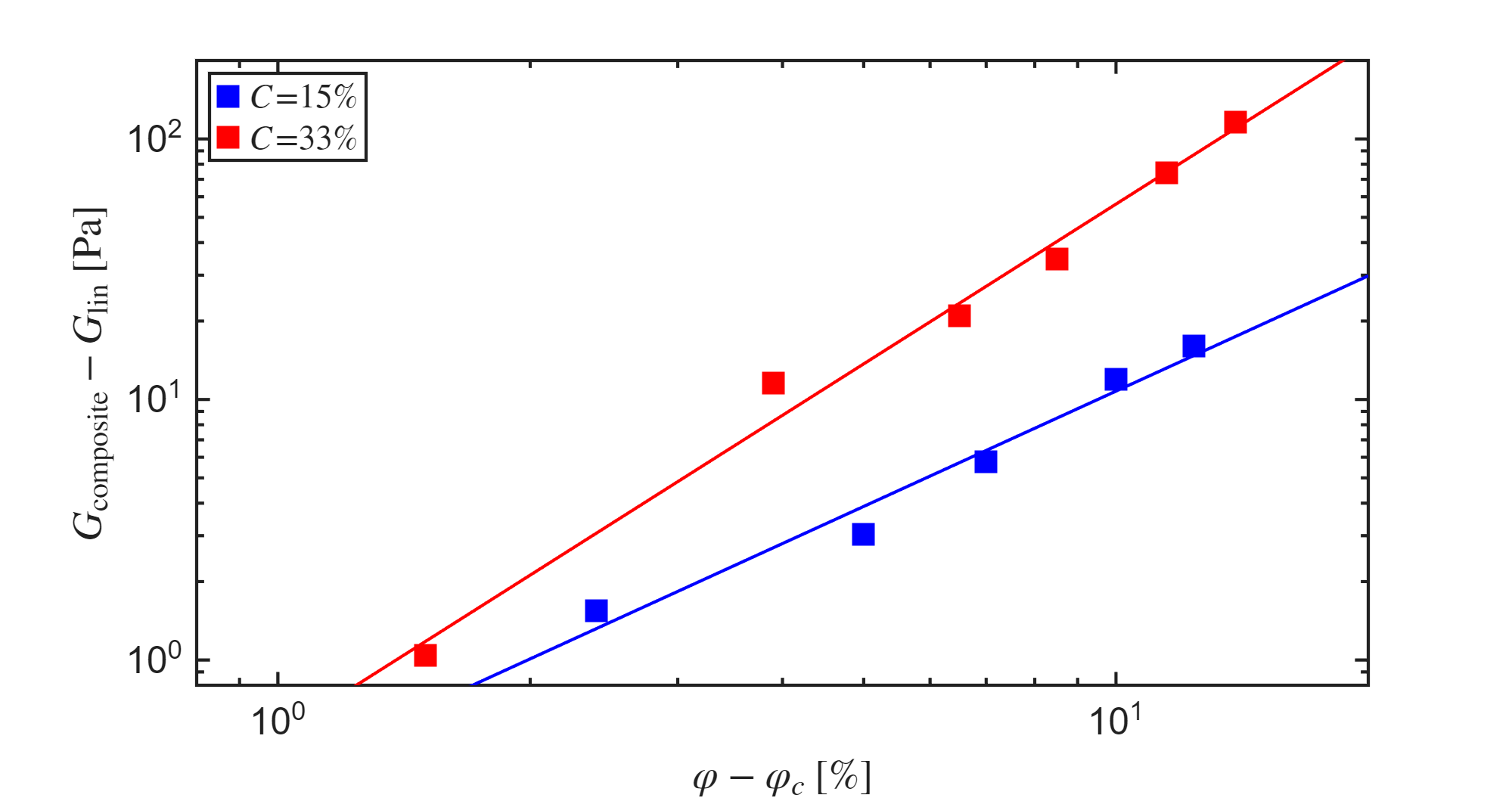}
\caption{Log-log plot showing that the deviation of the composite's shear modulus from its linear dependence on density scales as a power-law with the deviation of the density from the critical density.}
\label{fig:power-law}
\end{figure}

Below the critical density, $\varphi<\varphi_c$, the shear modulus of the composite agrees with the linear dependence expected by effective medium theory, $G_{\rm composite} \approx G_{\rm lin} = A + B \varphi$ with $A=13.8$, $B=0.586$ for non-contracting cells ($C=0$), $A=13.5$, $B=1.39$ for $C=15\%$ and $A=12.8$, $B=3$ for $C=33\%$ in nonlinear media, and $A=14.5$, $B=0.712$ for $C=33\%$ in a linear medium (dashed lines in Fig.~\ref{fig:material_properties}d). Above the critical density, $\varphi > \varphi_c$, we observe a sharp increase of $G_{\rm composite}$, and expect to find power-law scaling $G_{\rm composite} \propto (\varphi - \varphi_c)^f$ with $f>1$. As density increases, this power-law behavior dominates. However, to capture this power-law behavior with the data we have, even close to $\varphi_c$, we assume there are two additive effects causing the shear modulus to increase. Namely, that the linear dependence predicted by effective medium theory continues to contribute to the shear modulus also for $\varphi>\varphi_c$. In Fig.~\ref{fig:power-law} we show the agreement with this scaling, namely, $G_{\rm composite} - G_{\rm lin} \propto (\varphi - \varphi_c)^f$, with $\varphi_c=10\%$, $f=1.47$ for $C=15\%$ and $\varphi_c=8.5\%$, $f=2.04$ for $C=33\%$. 

\clearpage

\bibliography{bibliography}

\end{document}